\documentclass[useAMS,usenatbib,letterpaper]{mn2e}
\usepackage{times}
\usepackage{amssymb}
\usepackage{lscape,graphicx}

\title[The colour selection of distant galaxies in the UKIDSS Ultra-Deep Survey Early Data Release] {The colour selection of distant galaxies in the UKIDSS Ultra-Deep Survey Early Data Release}

\author [K. Lane et al.]
{K. P. Lane$^{1}$,
  O. Almaini$^{1}$, S. Foucaud$^{1}$, C.  Simpson$^{2}$, Ian
  Smail$^{3}$, R.~J. McLure$^{4}$, \newauthor C.~J.  Conselice$^{1}$,
  M. Cirasuolo$^{4}$, M.~J. Page$^{5}$, J. S. Dunlop$^{4}$,
  P. Hirst$^{6}$, M.G. Watson$^{7}$,\newauthor K. Sekiguchi$^{8}$ \\
  $^{1}$School of Physics and Astronomy, The University of Nottingham,
  University Park, Nottingham, NG7 2RD\\ $^{2}$Astrophysics Research
  Institute, Liverpool John Moores University, Twelve Quays House,
  Egerton Wharf, Birkenhead CH41 1LD\\ $^{3}$Institute for
  Computational Cosmology, Department of Physics, Durham University,
  Durham DH1 3LE\\ $^{4}$SUPA\thanks{Scottish Universities Physics
    Alliance} Institute for Astronomy, University of Edinburgh, Royal
  Observatory, Edinburgh EH9 3HJ\\ $^{5}$Mullard Space Science
  Laboratory, University College London, Holmbury St. Mary, Dorking,
  Surrey RH5 6NT\\  $^{6}$Gemini Observatory Northern Operations
  Center, 670 N. A'ohoku Place Hilo, Hawaii, 96720, USA\\ $^{7}$Department
  of Physics \& Astronomy, University of Leicester, Leicester LE1
  7RH\\ $^{8}$Subaru Telescope, National Astronomical Observatory of
  Japan, 650 North A'ohoku Place, Hilo, Hawaii 96720, USA}

\begin{document}

\date{}

\pagerange{\pageref{firstpage}--\pageref{lastpage}} \pubyear{}

\maketitle

\label{firstpage}

\begin{abstract}

We investigate colour selection techniques for high redshift galaxies in
the UKIDSS Ultra Deep Survey Early Data Release (UDS EDR). Combined with
very deep Subaru optical photometry, the depth ($K_{AB} = 22.5$) and area
($0.62$deg$^2$) of the UDS EDR allows us to investigate optical/near-IR
selection using a large sample of over 30,000 objects. By using the $B-z$,
$z-K$ colour-colour diagram (the BzK technique) we identify over 7500
candidate galaxies at $z>1.4$, which can be further separated into passive
and starforming systems (pBzK and sBzK respectively). Our unique sample
allows us to identify a new feature not previously seen in BzK diagrams,
consistent with the passively evolving track of early type galaxies at
$z<1.4$. We also compare the BzK technique with the $R-K$ colour selection
of Extremely Red Objects (EROs)  and the $J-K$ selection of Distant Red
Galaxies (DRGs), and quantify the overlap between these
populations. We find that the majority of DRGs, at these relatively bright
magnitudes are also EROs. Since previous studies have found that
DRGs at these magnitudes have redshifts of $z \sim 1$ we determine
that these DRG/ERO galaxies have SEDs consistent with being dusty
star-forming galaxies or AGN at $z<2$. Finally we observe a flattening in the number
counts of pBzK galaxies, similar to other studies, which may indicate that we are sampling the
luminosity function of passive $z>1$ galaxies over a narrow redshift
range.

\end{abstract}


\section{Introduction}

Colour selection can provide an efficient technique to identify
high-redshift galaxies based on known rest-frame SED
features. However, most of these colour criteria are sensitive to more than
one type of SED feature, i.e. more than one galaxy type and at potentially
different redshifts. 

EROs are one class of colour-selected galaxies, these are consistent with
either dust-reddened star forming galaxies or passive galaxies at $z > 1$
(e.g. \citealt{Cimatti_etal:2002}, \citealt{Roche_etal:2002},
\citealt{Simpson_etal:2006}). Extending these techniques to higher
redshifts \citet{Daddi_etal:2004} suggested a more refined criteria based on two
colour cuts in $B-z'$ and $z'-K$, now known as the BzK technique, where
$BzK\equiv(z'-K)_{AB}-(B-z')_{AB}$. Star forming galaxies in the range
$1.4 < z < 2.5$ are selected by requiring
$BzK\geq-0.2$. This was determined by analysing the BzK properties
of z $>$ 1.4 star-forming galaxies, classified through [OII] emission and UV
spectra. Passive galaxies in the same redshift range require $BzK < -0.2$ and
$(z'-K)_{AB} > 2.5$. This was determined by analysing the BzK properties of
spectroscopically confirmed passive galaxies at $z >
1.4$. Star-forming and passive BzK galaxies will now be referred to as sBzK
and pBzK respectively. Finally the selection of galaxies using near-infrared
colours $(J - K)_{AB} > 1.3$ (DRGs) has been shown to be an effective technique for
selecting galaxies out to $z > 2$ based on the Balmer break
\citep{Franx_et_al:2003}.

The effectiveness of these various photometric techniques, however, is still
unquantified, due in part to the small sample sizes, or shallow depths, used in
previous follow-up studies. The latest generation of large area, deep
near-infrared surveys, such as the UKIDSS UDS, now provide a representative sample of
colour-selected sources. These can be used to determine the proportions of
galaxy type selected using different colour criteria, their redshift ranges
and the overlaps between different techniques. 

Throughout this paper we use a concordance cosmology with $\Omega_m=0.3$, $\Omega_\Lambda=0.7$, and $H_0=70$
km s$^{-1}$ Mpc$^{-1}$. The magnitudes and colours quoted above are based on 2
arcsec diameter aperture magnitudes, as are all magnitudes and colours quoted throughout this
paper. 

\section{Colour Selection of High-z Galaxies}

\subsection{Data Set and Sample Definitions}
\label{BzK Selection}

The UKIRT Infrared Deep Sky Survey (UKIDSS) has been running since
spring 2005 and comprises 5 sub-surveys covering
different areas and depths \citep{Lawrence_etal:2006}. The UKIDSS survey uses the
Wide-Field Camera (Casali et al., in preparation) on the United Kingdom
Infrared Telescope (UKIRT). This study makes extensive use of $J$ and
$K$-band data from the Ultra Deep Survey early data release
(the UDS EDR; see \citealt{Dye_etal:2006}). The UDS is the
deepest of the 5 UKIDSS sub-surveys and aims to reach a final depth of
$K_{AB} =  25.0$, $H_{AB} =  25.4$, $J_{AB} =  26.0$ ($5\sigma$, point
source) over an area of 0.8 deg$^{2}$. The size of the UDS field significantly
reduces the effects of cosmic variance and on this scale the UDS will be the deepest
image ever produced, providing an unparalleled number of candidate
high-redshift sources. Due to a small change in the UDS field centre shortly after the beginning
of the survey, the current $JK$ imaging is not uniform over the entire
0.8 deg$^{2}$ field. Consequently, for the purposes of this paper, we
restrict our analysis to the 0.6 deg$^{2}$ central region which has uniform
$JK$ data and reaches depths of $K_{AB} =  22.5$ and $J_{AB} =  22.5$
($5\sigma$ limits). By conducting simulations, the completeness at this
magnitude limit is determined to be above 70\% for point sources. For
details of the completeness estimation, image stacking,
mosaicing and catalog extraction procedures see \citet{Foucaud_etal:2006}. In
addition to $J$ and $K$-band data from the UDS, deep $B$ ($B_{lim} = 28.2$), $V$
($V_{lim} = 27.5$), $R$ ($R_{lim} = 27.6$), $i$ ($i_{lim} = 27.5$) and $z'$-band
($z'_{lim} = 26.5$) data from Subaru Suprimecam are also used \citep{Sekiguchi_etal:2005}.

\begin{figure}

\includegraphics[width=0.99\columnwidth]{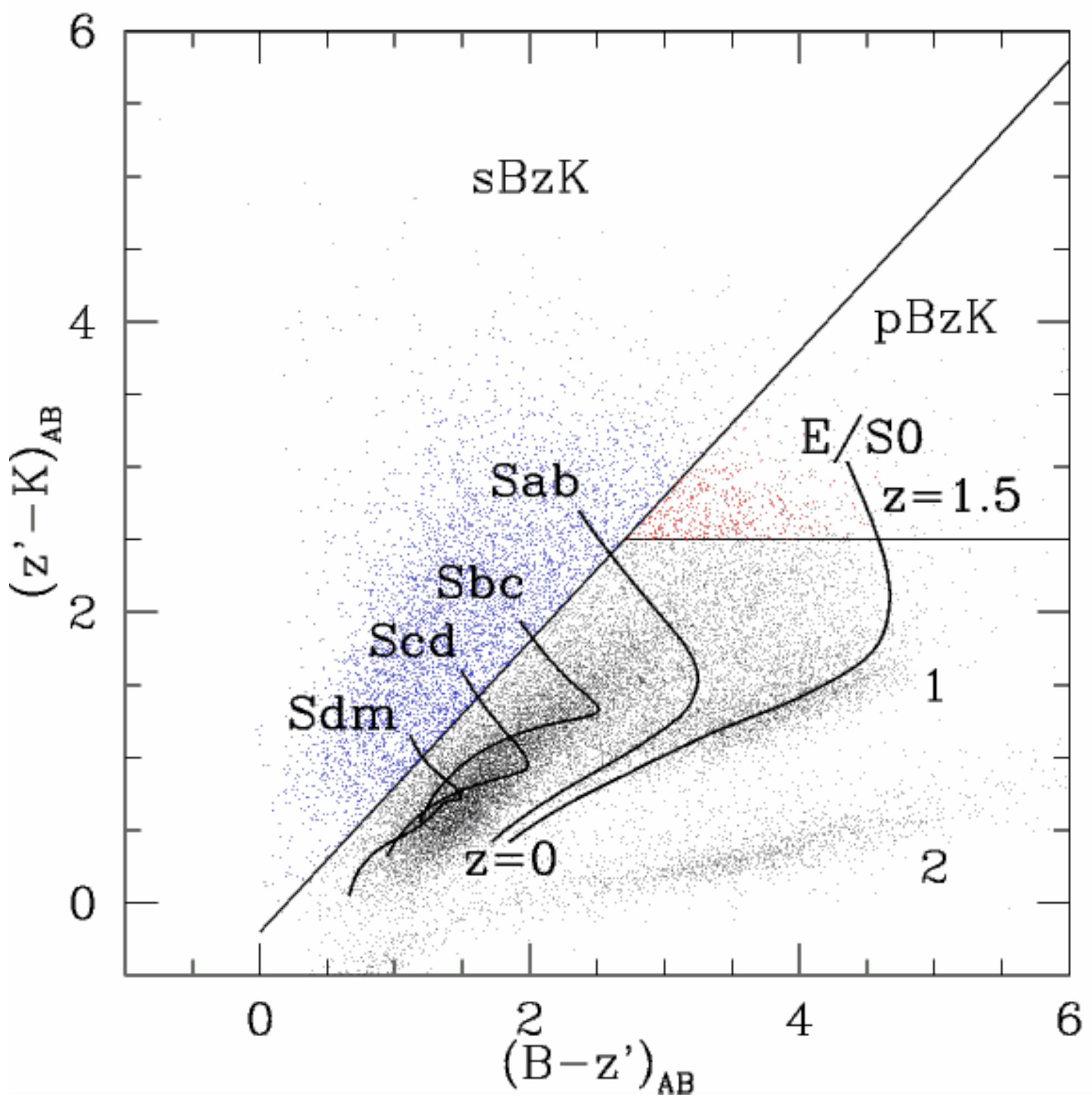}

\caption{Plot showing the BzK colour-colour diagram for sources in the UDS
  EDR. The sBzK and pBzK regions are marked accordingly. Branch 1 is the new
  branch of galaxies found in this study and Branch 2 is the stellar
  branch. Overlaid are no evolution model tracks for a range of SED types,
  starting at z$=$0 and ending at z$=$1.5.}

\label{fig-BzK}

\end{figure}

The wealth of multi-band data available in this field, together with their
depth and large area coverage, has enabled the construction of a highly-detailed BzK
colour-colour plot which displays many interesting features including a new
branch (Figure \ref{fig-BzK}). For consistency the standard BzK definitions of
\citet{Daddi_etal:2004}, as discussed in the introduction, were used to
construct our BzK sample. $K$ band sources ($>5\sigma$) were crossmatched with our
optical Subaru $i$ band source catalog; since the Subaru field is not
entirely coincident with the UDS field this reduced the usable
area. For $K$ band sources ($>5\sigma$) that did not have an optical
match, 2 arcsec aperture magnitudes were obtained directly from the
Subaru images. All BzK
sources were then visually verified to make sure they were not caused by
diffraction spikes from saturated sources or cross-talk effects
\citep{Dye_etal:2006}. Additionally, only sources that were not within the
halos of saturated optical sources were used, which left an area of
0.5591 deg$^{2}$. All $K$-matched sources were used for BzK selection
unless both the $B$ and $z$ band magnitudes were $<1\sigma$ limits
(0.7\% of the sample, outside of saturated halos), since these cannot be constrained within the
BzK plane. This resulted in the selection of 6736 sBzK and 816 pBzK.

In addition to BzK selection the availability of $R$ and $J$ band data in this field has also
enabled the extraction of EROs and DRGs. In this case
the ERO selection was defined as $(R - K)_{vega} > 5.3$ and is carried out
using the prescription presented in
\citet{Simpson_etal:2006}, resulting in 4621 EROs (to $K_{AB}=22.5$). ERO selection
was carried out independently because \citet{Simpson_etal:2006} uses a
different method of optical-NIR matching than used here. The DRG sample used here is constructed in
\citet{Foucaud_etal:2006} using a selection criteria of $(J - K)_{AB} > 1.3$,
resulting in 369 DRGs (to $K_{AB}=21.2$). Only those DRGs within the Subaru region
and outside of saturated halos are used, this leaves 330 DRGs suitable for
cross matching with NIR-optical sources.

\subsection{BzK Number Counts}
\label{Number Counts}

\begin{figure}

\includegraphics[width=0.98\columnwidth]{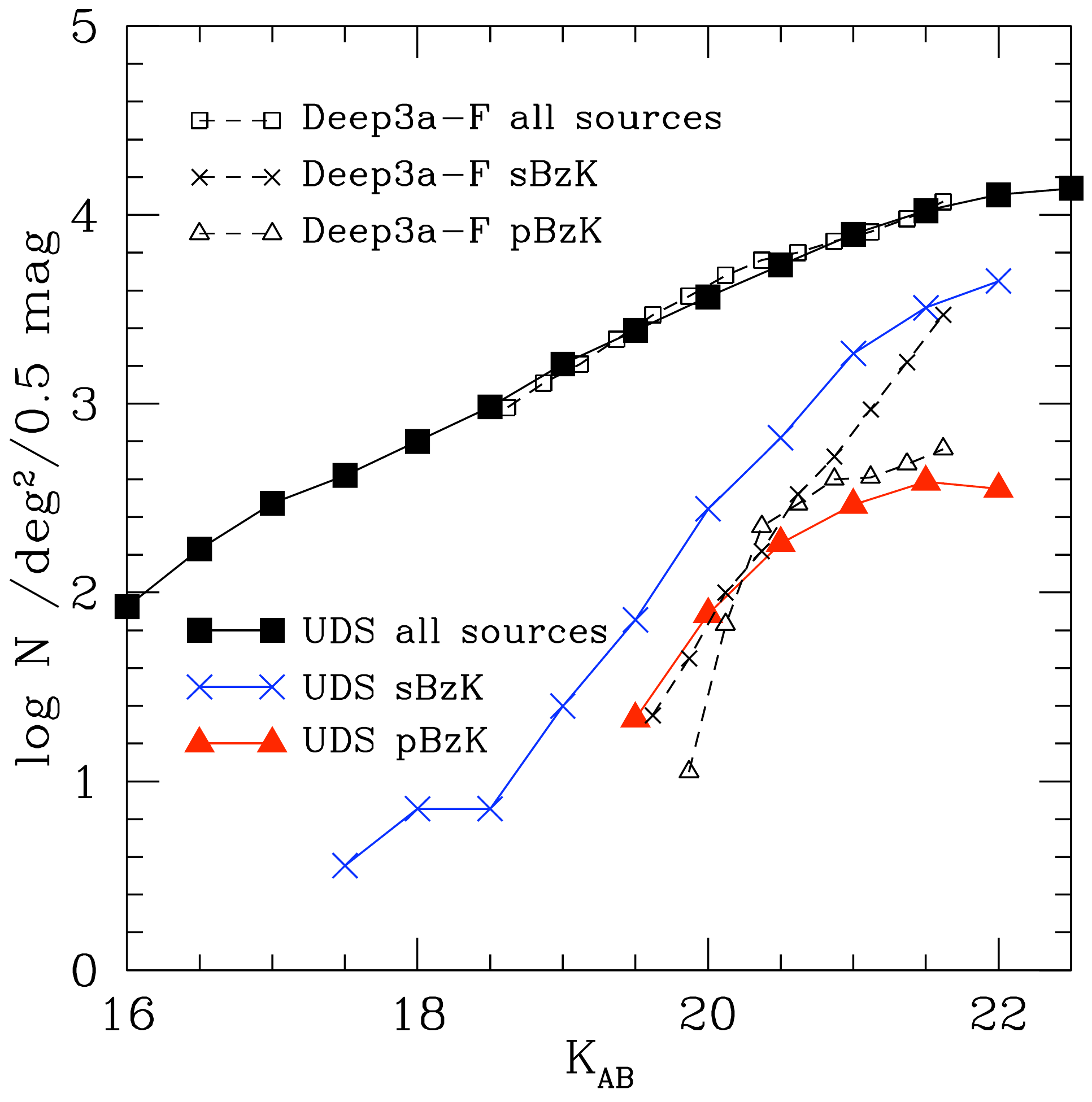}

\caption { K-band differential number counts for all sources in the field and
  BzK sources. The sBzK population rises rapidly to our completeness limit,
  while the pBzK counts exhibit a knee at $K_{AB}\sim21$. Data from the
  Deep3a-F sample of \citet{Kong_etal:2006} is included for comparison.} 
\label{fig-counts}

\end{figure}

The differential $K$-band number counts for the different populations in our
survey (Figure \ref{fig-counts}, Table \ref{tab-counts}) show that star-forming BzKs increase
in number sharply toward fainter magnitudes. In contrast pBzKs seem to exhibit a
knee at $K \sim 21$ where the number counts clearly turn over before the
limiting $K$-band magnitude, this feature is also seen in \citet{Kong_etal:2006}. The small redshift range over which pBzKs are
thought to be selected \citep{Kong_etal:2006} could provide an explanation for
the knee and the relatively faint magnitude at which pBzK number counts
begin. That is, that the number counts are directly sampling the
luminosity function of this population.

We note that the galaxies forming this knee feature have very faint
$B$-band magnitudes, close to the completeness limit in this band
($B_{lim}=28.2$). However fainter $B$ magnitudes will push the $B-z$
colours to the red and hence these sources should not be missed by our
pBzK selection. At $K_{AB} \sim 21$ completeness is $\sim 100\%$ \citep{Foucaud_etal:2006} so the
feature is unlikely to be due to incompleteness. We conclude that this knee feature is likely to be
real. We note that the turn-over corresponds to absolute magnitudes
$M_{K} \sim -23.6$ at $z=1.4$, which is close to the value of $M^*$
determined for the $K$-band luminosity function at these redshifts \citep{Cirasuolo_etal:2006}. 

As can be seen in Figure \ref{fig-counts} our results are similar in form to
the data from the Deep3a-F survey \citep{Kong_etal:2006}, especially for the overall source number counts. However, for the sBzK number counts there is a slight magnitude offset, most likely due to cosmic variance. The UDS EDR has
greater dynamic range in $K$ due to our greater depth and area. This
dynamic range will increase toward even fainter magnitudes with further UDS releases.

\begin{table}
\caption{Differential Number Counts in log(N/deg$^{2}$/0.5mag) bins
  for sBzK and pBzK in the UDS EDR.}
\begin{tabular} {|l|l|l|l|}
\hline
K bin centre&all sources&sBzK&pBzK\\
\hline
16&1.925&-&-\\
16.5&2.23&-&-\\
17&2.473&-&-\\
17.5&2.62&0.554&-\\
18&2.8&0.855&-\\
18.5&2.983&0.855&-\\
19&3.21&1.399&-\\
19.5&3.388&1.855&1.332\\
20&3.565&2.443&1.886\\
20.5&3.734&2.82&2.261\\
21&3.897&3.266&2.465\\
21.5&4.022&3.508&2.587\\
22&4.107&3.65&2.549\\
\hline

\end{tabular}
\label{tab-counts}
\end{table}

\subsection{BzK New Branch Galaxies}
\label{NewBranch}

Due to the large number of detected sources in this survey field (34098 in total), and the comparative depth of
the optical data available, there are a large number of sources available for
BzK analysis. It is clear from Figure \ref{fig-BzK} that a new feature is
visible running parallel with the stellar branch, at the bottom of the plot,
but with redder $z'-K$ colour, roughly defined by the region: \begin{equation} 0.3 (B-z')
- 0.2 \lesssim (z'-K) \lesssim 0.3 (B-z') + 0.4  \end{equation}  with clear
branch separation at $(B-z') \gtrsim 2.5$.

Also plotted on Figure \ref{fig-BzK} are no-evolution model tracks for
different types of galaxy. These were created by redshifting SEDs from
\citet{King_Ellis:1985}, convolved with the appropriate filter response
curve to get the relevant band magnitudes from which the $(B-z')$ and $(z'-K)$ colours
can be calculated. The predicted
track for early type galaxies (E/S0)
corresponds very well with the new branch of galaxies, especially at
low redshifts. As can be seen, at higher redshift galaxies are no
longer populating the predicted E/S0 model track. This is partly due
to incompleteness at these high redshifts but also because these model tracks do not take into account
evolution, so early type galaxies will be bluer at higher redshifts than the
models predict. As such they only provide a guide as to where
different types of galaxies exist within the BzK plane. The tracks of later-type spirals fall
predominantly within the main trunk of galaxies in the BzK diagram and all lie
within the sBzK region by z$\sim$1.5. The proximity of the Sab model track to
that of E/S0 suggests that there may be some contamination of the
early type branch by spirals, especially at low redshift. 

\section{ERO/BzK comparison}
\label{ERO/BzK}

The large number of EROs, BzKs and DRGs provided by the UDS EDR for the first
time allows the overlaps between these populations to be analysed in
detail. Previous studies have been made of the overlaps between EROs
and BzKs, using smaller numbers of sources than are available
here. \citet{Kong_etal:2006} find that 41$\%$ of EROs are also selected as
BzKs in their Deep3a-F data, and 29$\%$ from their Daddi-F data. This compares
with \citet{Daddi_etal:2004} who find that $\sim$35$\%$ of their ERO selected
sources are also BzK selected. In our much larger study we find that $60.6\pm1.5\%$
of EROs (to $K_{AB}=22.5$) are
selected as BzK, of either variety, which is a higher fraction than found in previous
studies. This is not surprising since the UDS EDR is deeper and contains fainter
EROs, of which a larger fraction are likely to represent $z>1.4$
objects than would be expected in a brighter sample.

\begin{figure}

\includegraphics[width=0.98\columnwidth]{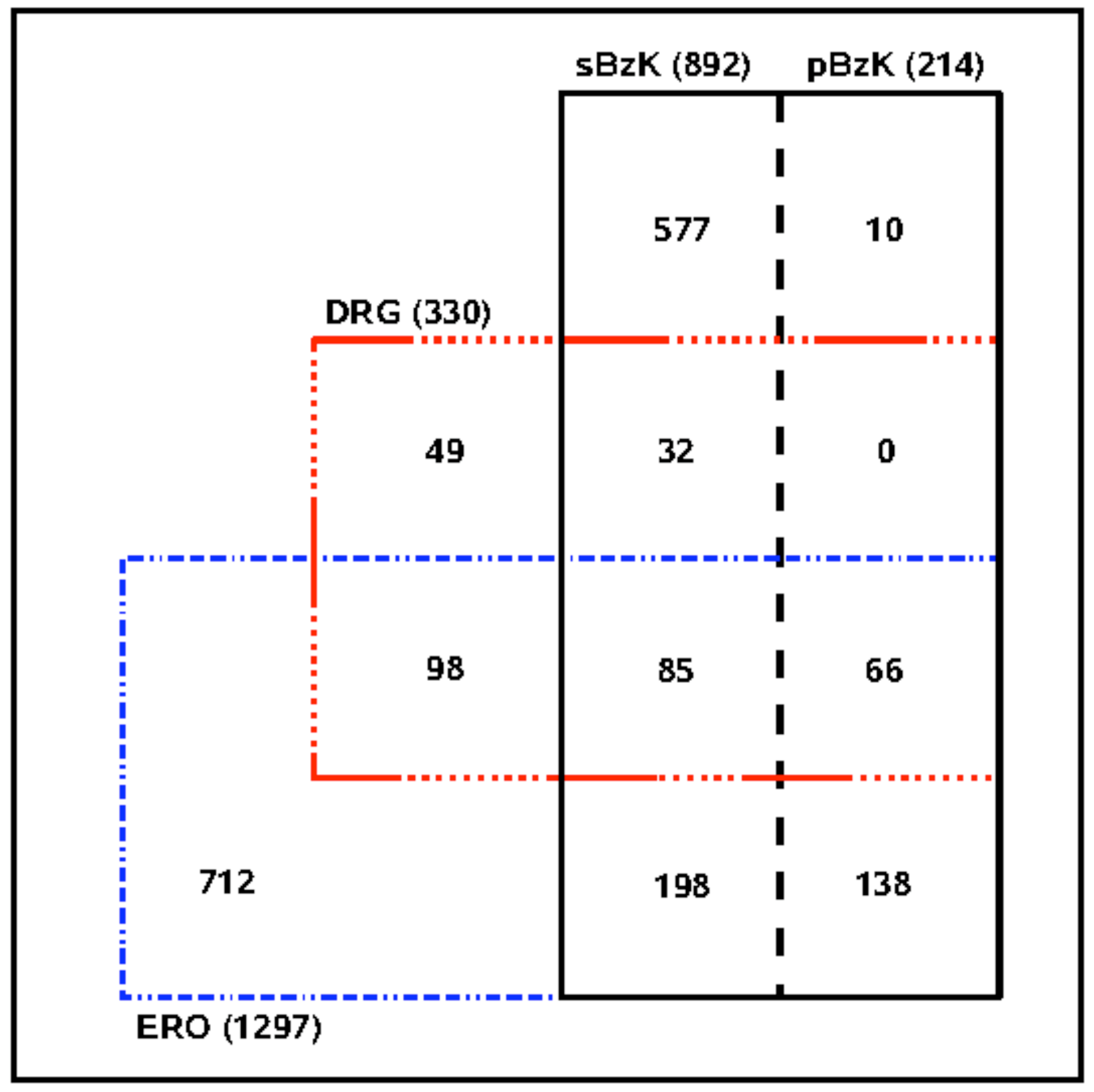}

\caption{Venn diagram showing the overlaps in the number of objects classified
  as EROs, BzKs and DRGs. All numbers given here are to the depth of the DRG
  sample ($K_{AB}=21.2$).} 
\label{fig-Venn}

\end{figure}

\begin{figure}

\includegraphics[width=0.97\columnwidth]{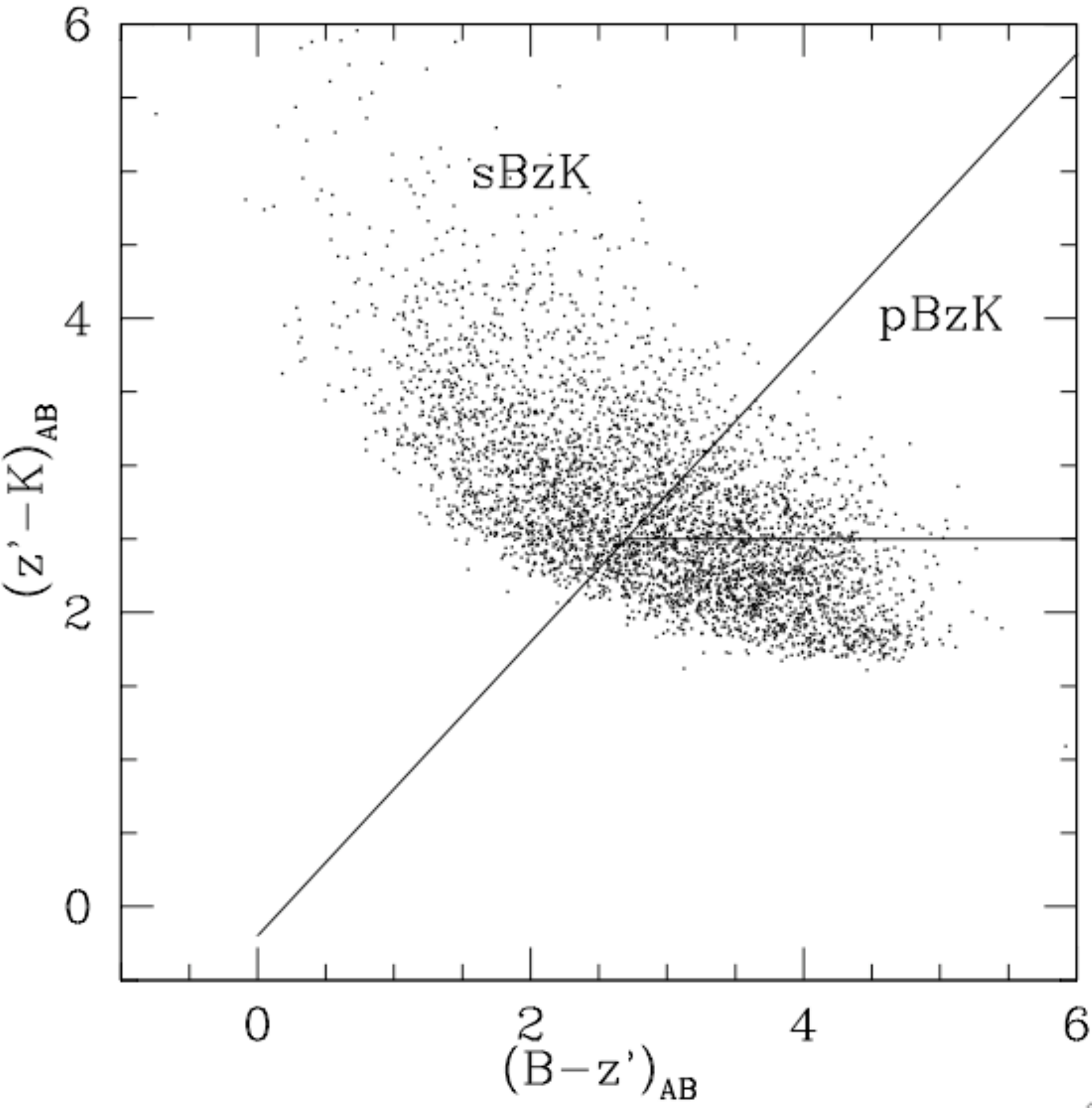}

\caption{ Figure showing the BzK colours of EROs. Almost all pBzKs are
  EROs as well as a large fraction of sBzKs. This can also be seen in
  Figure \ref{fig-Venn}. } 
\label{fig-erobzk}

\end{figure}

Figure \ref{fig-Venn} shows the overlaps between
the four populations studied here and the number of sources involved in
each case. To ensure we are only comparing samples to the same completeness
depth only EROs and BzKs selected to $K_{AB}=21.2$
are used in this diagram, since this is the limit for DRG
selection. However, when comparing EROs and BzKs we also quote numbers
to a depth of $K_{AB}=22.5$ in parentheses in the following
discussion. As can be seen, 204(792) out of 214(816)
pBzKs are additionally selected as EROs. The reason for this strong
overlap is clear from the position of EROs on the BzK
diagram (Figure \ref{fig-erobzk}); EROs are located in a clearly defined region
with red colours in $z'-K$ and $B-z'$ which covers the entire region occupied by pBzKs. It is also clear from Figure
\ref{fig-erobzk} that sBzKs are far less preferentially selected as EROs -
283(2007) out of 892(6736) sBzKs, or $31.7\pm1.4\%$($29.8\pm1.2\%$). 

\begin{figure}

\includegraphics[width=0.98\columnwidth]{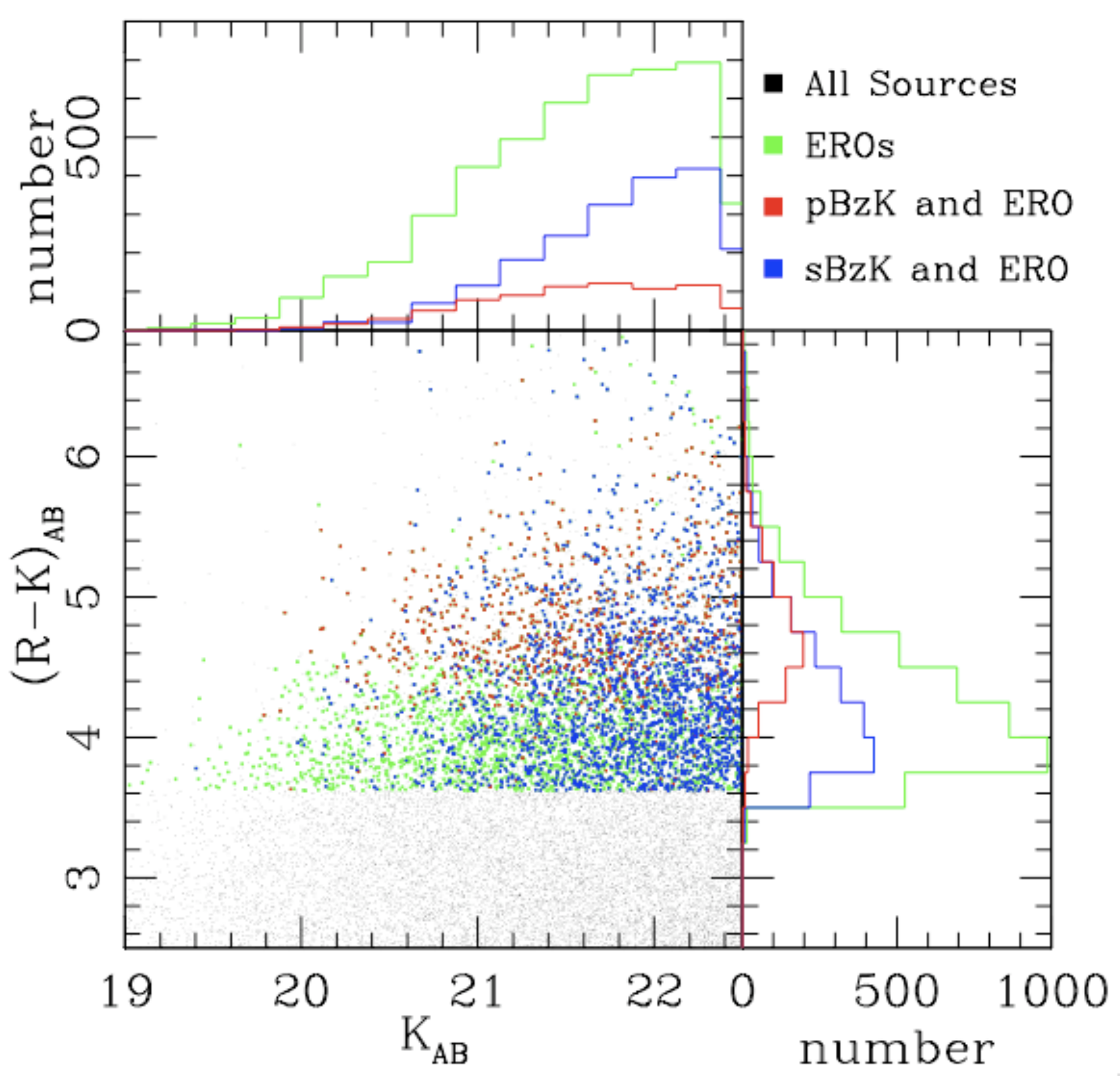}

\caption{ An $R-K$ diagram showing BzK selected sources that are also selected
  as EROs. The histogram to the right shows the $R-K$ colour of EROs and ERO
  selected p and sBzK. It is clear that pBzKs have
  redder colours, whereas sBzKs cover a broad range in $R-K$ colour. The top
  histogram shows $K$-band aperture magnitudes for the same subsets. Much like
  the overall ERO population, sBzKs tend
  to be at higher magnitudes and peak around $K_{AB}\sim22$, conversely pBzKs have a much
  broader distribution which peaks at $K_{AB}\sim21.2$.} 
\label{fig-bzkRK}

\end{figure}

From the position of BzKs on a plot
of $R-K$ colour versus $K$ magnitude (Figure \ref{fig-bzkRK}) it can be seen that
there is a clear difference in distribution  between pBzKs and sBzKs. pBzKs have redder $R-K$
colour than sBzKs, with a narrow range around $R-K = 4.6$ (FWHM$\sim$0.24, see side
panel in Figure \ref{fig-bzkRK}). This is in contrast to sBzKs which tend to be broad in
$R-K$ colour but occupy regions with fainter $K$ magnitude (see top
panel in Figure \ref{fig-bzkRK}). This could be explained if sBzKs constitute
the higher redshift ($z>1.4$) end of the ERO
population. Using the photo-z catalog presented in \citet{Cirasuolo_etal:2006}
shows the same redshift behaviour. However, given that these photo-z values are based on the same
photometric data used here to construct the BzK population, and use
the same SED features used to place galaxies at $z \sim 1.4$ this does not
essentially add any new information, but serves as a consistency check.

\section{DRG/ERO comparison}
\label{DRG/ERO}

The DRG selection technique was originally intended to preferentially select
galaxies at $z>2$. However the most recent generation of large area NIR
surveys (e.g.  \citealt{Grazian_etal:2006},
\citealt{Conselice_etal:2006}) are starting to probe the bright, low
redshift, end of the DRG population. Based on photometric redshifts \citet{Quadri_etal:2007} find $>70\%$
of DRGs (to $K_{vega} < 20$) from the Multiwavelength Survey by
Yale-Chile (MUSYC) to lie at $z>1.8$. Compared to the UDS EDR, however, the
greater depth and substantially smaller area used in
\citet{Quadri_etal:2007} means that a much higher fraction of the
DRGs selected are likely to lie at higher redshifts. A more
comparable study is \citet{Conselice_etal:2006} in which a DRG
sample is produced to a depth of $K_{vega}=20.5$ over a slightly larger
area than used in this study. They find their DRG sample to peak at $z\sim1.2$, when using photometric
redshifts, and at $z\sim1.0$, when using spectroscopic redshifts, with
only 4\% at $z>2$. This redshift distribution does not change when
their DRG sample is cut to the same depth as the DRG sample
constructed in this study ($K_{vega}=19.3$). We therefore take $z\sim1.0$ to also be the likely redshift
distribution of our DRG sample. A new picture is starting to emerge in which the DRG
criterion not only selects $z>2$ passive galaxies, but also galaxies at
redshifts of $z\sim1$. It is therefore interesting to see how DRGs cross match
with photometric populations of a more established nature, such as EROs and
BzKs.

Most DRGs in the UDS EDR are also selected as EROs ($75.5\pm6.3\%$, see Figure \ref{fig-Venn}). This means
that these galaxies are red in both $R-K$ and $J-K$ colours which implies
they have a steep slope in their SEDs across this colour range. An SED
feature of this kind, at $z\sim1$, is indicative of dusty star-forming galaxies or
AGN. {\it XMM-Newton} X-ray data is available for this field (Ueda et
al. in prep.), but only 6 ($1.8\pm0.7\%$) of our DRGs
are found to be X-ray detected. At the depth of the XMM data, however, this is only sufficient
to rule out luminous AGN with relatively modest $N_H$ columns
(e.g. $N_H<10^{23}$cm$^{-2}$, assuming $L_X \simeq 10^{44}$ erg s$^{-1}$ at $z=1.5$). Therefore the presence of
highly obscured and/or lower luminosity AGN cannot be ruled
out. Of the joint ERO and DRG galaxies, $34.1\pm4.3\%$ are sBzK and $26.5\pm3.7\%$ are pBzK. The number of
pBzKs in this overlap group suggests that a sizeable portion of these pBzKs are
likely to be either obscured AGN or dusty star forming galaxies rather than
purely ``passive'' pBzK. 

It would therefore appear that DRGs not only select $z>2$ galaxies but
also form part of the z$\sim$1 ERO population, based in part on the
DRG redshift determination from \citet{Conselice_etal:2006}, as dust reddened star
forming galaxies as well as a fraction of
possible obscured AGN (see also \citealt{Conselice_etal:2006}). 

\section{DRG/BzK comparison}
\label{DRG/BzK}

Approximately half of the galaxies selected as DRGs are also selected as
BzKs (117 sBzK and 66 pBzK). Although DRGs are selected as BzKs across the full DRG $K$ magnitude and $J-K$
colour range, they tend to have fainter $K$ magnitudes and
are more efficiently selected by redder $J-K$ colours. Of these joint
selections, $72.6\pm10.4\%$ of the sBzK selected DRGs were also selected as EROs, as
were all of the pBzK selected DRGs. This is as expected, due to the large fraction
of DRGs selected as EROs (section \ref{DRG/ERO}).  

\citet{Reddy_etal:2005} find that $\sim$30\% of DRGs are found to be sBzKs, which is similar to the
fraction of DRGs found as sBzKs in our study. The study also finds
that $\sim$10\% of sBzKs are DRGs, as are $\sim$30\% of
pBzKs; though sample errors will be high since only 17
pBzKs are used. This is in good agreement with this paper which finds that
$13.1\pm1.3\%$ of sBzKs and $30.8\pm4.3\%$ of pBzKs are also selected as DRGs. The large area
of the UDS EDR data is probing the lower
redshift, brighter end of the DRG luminosity function, so a large overlap
with BzK-selected galaxies would be expected (we observe $\sim$55\%). 

The larger fraction of pBzKs selected as DRGs could be due to the narrower and
lower redshift range of pBzKs, as
hypothesised in previous studies, combined with the DRG technique selecting
galaxies at z $\sim$ 1. It could also be due to some shared
astrophysical feature, such as AGN (as discussed in section
\ref{DRG/ERO}). Whichever is correct, it is likely
that BzKs constitute at least the $z \sim 1.4 - 2.0$ part of the DRG
population. Those DRG not selected as BzK are likely to be at $z <
1.4$, as in \citet{Daddi_etal:2004}, though some could be at $z > 2.5$.

\section{Conclusions}
\label{Conc}

For the first time a statistically significant study of the overlaps between
ERO, BzK and DRG galaxy populations has been carried out. We compare 1297
EROs, 330 DRGs and 1106 BzKs, selected from 0.5591 deg$^{2}$ of imaging
to $K_{AB} = 21.2$. It is found that
BzKs are consistent with being the  $z > 1.4$ end of the ERO population, as would be
expected from the definition of these selection techniques.

It is becoming clear from the new generation of large area surveys that the
DRG selection criterion is not only effective at
extracting high z galaxies, but also dusty star forming galaxies and
obscured AGN, at z$\sim$1, particularly at brighter $K$-band
magnitudes ($K_{AB} < 22$). In this study we find that 249 of 330 DRGs are also
selected as EROs. Those DRGs also selected as pBzKs tend to be at the red
end of the DRG sample while those also selected as sBzKs have a range in
$J-K$ colour but are faint in
$K$-band magnitude. This is consistent with DRGs selecting galaxies over a broad
redshift range, from EROs at $z \sim 1$ to BzKs at $z \sim
1.5$. Deeper UDS data will also allow us to probe the faint end of the
DRG selection regime to ascertain the effectiveness of this technique at
selecting galaxies at $z > 2$.

The depth of our Subaru optical data has allowed us to determine the
number-magnitude relations of the BzK samples. sBzKs have a broad
range in magnitude, consistent with a broad range in redshift and/or
luminosity. The pBzKs, however, exhibit a turn over in their number counts
that are consistent with pBzKs inhabiting a narrow redshift range at $z \sim
1.5$. The UDS EDR combined with this Subaru data has also allowed the construction
of the most complete BzK diagram yet seen. The number of sources has allowed
the identification of a new feature that is most likely to be the
no-evolution track of passive early type galaxies with increasing redshift.

\section*{Acknowledgements}

This work is based partly on data obtained as part of the UKIRT Infrared Deep
Sky Survey. We are grateful to the staff at UKIRT for making these
observations possible. We also acknowledge the Cambridge Astronomical Survey
Unit and the Wide Field Astronomy Unit in Edinburgh for processing the UKIDSS
data. KPL, SF and CS acknowledge funding from PPARC. OA, IS and RJM acknowledge
the support of the Royal Society. We thank the anonymous referee for comments which greatly
improved the reliability of the results presented.

\bibliographystyle{mn2e.bst}
\bibliography{mn-jour,papers_cited_by_KL}

\label{lastpage}

\end{document}